# From Papers to Interpretive Knowledge Nodes: Proposing the Missing Object in Scholarly Knowledge Circulation

***Note:*** *(Interpretive Knowledge Nodes Series — Paper 1: Theoretical Foundations)*

**Li Li[1],Yu Cao[2]**

[1]Hefei No.62 Middle School, Hefei 230031,Anhui, China
[2]Anhui NARI ZT Electric Co., Ltd., Hefei 230031, Anhui,China

**Abstract**

The modern scholarly communication system, with the paper at its core, has successfully solidified "research outputs" into citable and traceable scholarly objects. Yet, across the full chain from knowledge production to knowledge reuse, a critical link — interpretation — has long existed without ever being objectified. The theoretical elaborations, methodological translations, and conceptual clarifications that researchers perform when reading papers constitute the factual foundation of knowledge circulation. However, these interpretive activities remain tethered to individual competence and informal communication channels, lacking an independent scholarly identity. The central question of this paper is: Can interpretive knowledge qualify as an independent scholarly object? We argue that the Interpretive Knowledge Node (IKN), while manifesting as a "node" within knowledge networks, should be defined in its object identity as a new kind of scholarly object. Drawing on scholarly object theory, this paper identifies a structural gap in the existing system — namely, that "interpretive knowledge formed after the process of understanding" lacks an objectified identity. By distinguishing the essential difference between "interpretation" and "information extraction," it establishes the cognitive boundary of the IKN. Finally, it defines the four essential attributes that an IKN must possess as a scholarly object: provenance, interpretation fixation, human recognition, and citability. The argument demonstrates that the IKN is not a replacement for the paper system but an expansion of the boundary of the scholarly object system — marking a critical step from merely recording "what has been discovered" toward simultaneously recording "how to understand those discoveries."

## 1 Introduction

### *1.1 The Paper: The Historical Success of the Scholarly Object*

The modern system of scientific communication rests upon an architectural foundation that is at once simple and remarkably effective: research produces papers, papers interconnect through citations, forming a traceable knowledge network. The core component of this architecture — the scholarly paper — is not merely a genre of writing but a mature and pervasive scholarly object[1][2].

As a scholarly object, the paper accomplishes a critical institutional task: it enables "research outputs" to exist independently of the researcher who produced them[3]. This achievement is the result of multiple converging infrastructures. The printing press solved the problem of mass reproduction, allowing knowledge to transcend spatial and temporal boundaries. The academic journal system introduced editorial selection and periodical publication, endowing research outputs with temporality and predictability. The peer review mechanism conferred academic credibility upon papers, transforming them from personal claims by their authors into "certified knowledge" scrutinized by the scholarly community. The DOI (Digital Object Identifier) system assigned each paper a unique and persistent identity, enabling stable locatability within the digital network[4]. Citation conventions embedded papers into traceable intellectual genealogies, making each paper both an inheritor of prior knowledge and a point of departure for subsequent research[5].

Thus, the paper does not merely carry knowledge — it is itself the most fundamental organizing unit of the scholarly system. It is the basic unit of knowledge recording — researchers' findings enter the public domain with the paper as their container[6]. It is the basic unit of citation — all scholarly acknowledgments and attributions of priority ultimately point to papers[7]. It is also the basic unit of evaluation — bibliometric analyses for hiring, promotion, and resource allocation take the paper as their smallest analytical unit[8]. This tripartite role makes the paper the first, and most mature, scholarly object in the modern academic system.

*1.2 The Gap in Knowledge Flow: From "Node" Discovery to "Object" Inquiry*

Yet the very success of the paper as a unit of knowledge organization has obscured a structural gap in the chain of knowledge flow.

In the first paper of this series[9], we revealed a dimension long neglected within citation networks — the Citation Pathway. We demonstrated that knowledge does not travel directly from original sources to ultimate citers but instead flows through a series of intermediate nodes. Among these, the Interpretive Knowledge Node (IKN) performs the critical function of transforming core concepts, methods, and argumentative structures from classic works into standardized, citable forms. The first paper demonstrated, from a scientometric perspective, the structural function that IKNs serve as "nodes" within citation networks: by compressing citation pathways and carrying interpretive activity, they shape the network topology of knowledge dissemination.

However, the first paper's argument stopped at the level of network-level functional description, leaving a more fundamental question unanswered: If interpretive knowledge indeed performs an irreplaceable function in knowledge dissemination, should it, like the "papers" it mediates, acquire an independent "object" identity?

This question points toward an issue at the level of object identity, not merely network structure. In the current academic system, interpretive activities — whether a researcher's theoretical elaboration of a classic text, a judgment about the applicability of a method, or the clarification and extension of a concept — have always existed as appendages: attached to specific paper texts (such as a paragraph in a review), attached to specific individuals (such as a professor's classroom teaching), attached to specific informal channels (such as academic blogs or conference discussions). Interpretive activity itself has never acquired an independent scholarly object identity — it has no independent identifier, no fixed citable form, no institutionalized certification mechanism.

In other words, the first paper told us: IKNs do something important within the network. This paper asks: Do IKNs themselves qualify as independent entities?

### *1.3 The Core Question: Does Interpretive Activity Qualify as a "Scholarly Object"?*

This inquiry is far from a purely metaphysical exercise — it carries profound institutional implications.

The current inventory of scholarly objects runs roughly as follows: papers (recording research outputs), datasets (providing data materials), software (implementing computational functions), review articles (synthesizing domain progress). Together, these objects constitute the infrastructure of scholarly knowledge. Yet a closer examination of this inventory reveals a conspicuous structural absence: the current system records "what research has produced," but it does not record "how researchers understand these productions." Papers record research findings, but they do not independently record the interpretive understanding of those findings. Review articles contain interpretation, but that interpretation is encapsulated within the author's personal viewpoint rather than existing as an independently citable, independently certifiable, independently updatable and iterable object.

The consequences of interpretation's lack of objectified identity are manifold: it is untraceable (one cannot determine the origin and dissemination path of a particular interpretation), uncertifiable (one cannot judge whether an interpretation has undergone communal scrutiny and endorsement), uncitable (one cannot directly cite "Person X's interpretation of Concept Y" in a reference list as an independent evidentiary basis), and non-cumulative (interpretation follows the person; it cannot form iterable knowledge assets).

The core research question of this paper is therefore: Does the scholarly system need a new kind of scholarly object to formally carry "interpretive knowledge"? And does the IKN qualify as the concrete form of this new object?

### *1.4 Contributions of This Paper*

This paper aims to accomplish three core tasks:

First, identify the object gap. Through a systematic analysis of the existing structure of the scholarly object system, it pinpoints the structural void where "interpretive knowledge units" have long existed but lack object identity.

Second, argue for object identity. It argues that the IKN is not merely a network node (its positional function within the knowledge network) but should be defined as an independent scholarly object (its object identity).

Third, define object attributes. It articulates the four essential attributes — provenance, interpretation fixation, human recognition, and citability — that an IKN must possess as a scholarly object, providing a conceptual foundation for its subsequent institutional realization.

## 2 The Paper as the Dominant Scholarly Object: A Historical Success

Before arguing why the IKN should be regarded as a new scholarly object, it is necessary first to examine the formative logic of the existing scholarly object system — in particular, how the paper became the dominant scholarly object. This historical review is not mere background exposition but provides a critical reference frame for the subsequent argument: understanding how the paper became a scholarly object helps us judge whether the IKN satisfies the same conditions for "objectification."

*2.1 The Historical Logic of the Paper Becoming the Core Object*

The establishment of the paper as a scholarly object was not the result of a one-time institutional design but rather the product of multiple technological infrastructures and social institutions gradually coupling over centuries.

The printing press was the first critical condition. Before the print era, knowledge dissemination depended on manuscripts, each copy unique and costly. The Gutenberg revolution made the mass reproduction of texts possible, providing scholarly papers with the material basis for dissemination across geographic boundaries. The academic journal system further introduced mechanisms of periodical publication and editorial selection — the Philosophical Transactions of the Royal Society, founded in 1665, marked the moment when scholarly outputs began entering the public domain in a standardized, periodic, and predictable form[10]. Peer review, gradually institutionalized in the mid-20th century, conferred upon papers the function of "scholarly certification": a paper was no longer merely its author's claim but "qualified knowledge" that had undergone communal scrutiny[11].

Entering the digital era, the emergence of the DOI system completed the final step in the paper's objectification. The DOI assigned each paper a unique, persistent identifier, making the paper a precisely locatable and persistently citable entity within the digital network[4]. The importance of this mechanism is often underestimated: for an entity to become a "scholarly object," it requires not only content but also a stable identity anchor — an identification mechanism that enables other researchers to say, "I am citing this, not that."

Finally, the institutionalization of citation conventions — from footnotes to reference lists, from unstructured text to structured metadata — made the relational links between papers traceable and measurable. A paper was no longer an isolated text but a node embedded within a knowledge network, its position and connections defining its role in the academic ecosystem.

*2.2 What the Paper Model Achieved*

In sum, the core achievement of the paper model lies in successfully realizing the objectification of research outputs. "Objectification" refers to the transformation of something originally dependent on a specific person or context into a stable, examinable, archivable, and independently citable entity[3]. The social function of this transformation is far-reaching:

- Detachment from the author: The scholarly efficacy of a paper does not depend on the author's personal presence or explanation. Other researchers can independently read, evaluate, and use its content[3].
- Temporal stability: Once published and assigned an identifier, the paper's content and identity are fixed. Subsequent errata, revisions, or retractions are operations performed upon an existing object, not negations of its existence.
- Examinability: Peer review and post-publication scholarly debate are possible precisely because the paper, as a stable object, can be repeatedly scrutinized.
- Connectability: Papers form a traversable knowledge network through citing and being cited[7]. A paper's value derives not only from its own content but also from its position within this network.

These achievements make the paper not merely a textual genre but an institutional object — one embedded within the core institutions of scholarly credit allocation, priority determination, and academic evaluation, serving as the underlying infrastructure for their operation.

*2.3 The Boundaries of the Paper Model*

Precisely because of the paper model's enormous success, however, the scholarly community has tended to overlook its boundaries — mistaking the functions the paper can perform for the functions it can perform for all knowledge organization purposes.

The first boundary is granularity mismatch. The paper takes the "complete study" as its recording unit. A typical paper contains a literature review, theoretical framework, methodological design, experimental analysis, discussion, and outlook — among other components. Yet when subsequent researchers "use" the knowledge from a paper, they are often not calling upon the entire paper but upon a specific knowledge unit within it — a concept's definition, a method's parameter settings, an experiment's numerical result, an intermediate conclusion from a theoretical derivation. A structural gap exists between the "whole" granularity of the paper and the "unit" granularity of knowledge reuse[8].

The second boundary is dimensional absence. The paper records "what the researcher produced" — what they discovered, verified, proposed — but it does not independently record "what the researcher understood." When a researcher reads a classic text and forms their own understanding of its meaning, scope of applicability, and limitations, this understanding is either assimilated into the researcher's own new paper (becoming part of that paper's argument rather than an independent knowledge object) or remains in private notes or informal exchanges. The paper system excels at recording production results but neglects recording understanding results[3].

The conclusion that follows is this: the paper is the best object for recording research outputs, but it is not the only object — and certainly not a sufficient object — for recording knowledge understanding. This conclusion lays the foundation for the argument below that the IKN is necessary as a new kind of scholarly object.

## 3 The Missing Scholarly Identity of Interpretive Activity

### *3.1 Interpretation Is Unavoidable in Knowledge Flow*

Within the complete chain from scholarly knowledge production to reuse, there exists a link that appears ordinary yet is profoundly consequential: interpretation[2]. "Interpretation" here refers not to "explanation" or "clarification" in the everyday sense but to a cognitive operation in the scholarly epistemological sense: the transformation of existing knowledge into a form that can be understood and mobilized within new contexts.

Interpretive activity is ubiquitous in scholarly practice and takes at least the following forms:

- Theoretical elaboration: The contemporary reading of a classic theory. For example, Darwin's theory of natural selection has been elaborated in different ways by different schools across different eras — Neo-Darwinism, the neutral theory, punctuated equilibrium — each elaboration going beyond "reiterating the original text" to endow the theory with new contextual meaning and scope of application.
- Methodological translation: Taking a method developed in one discipline and, through reinterpretation of its principles, translating it into another. The backpropagation algorithm of deep learning was derived from automatic differentiation theory and, after interpretive exposition, entered neuroscience, materials science, and financial modeling — each translation relying on an interpretive reconstruction of the original method, not a mechanical reiteration of the original paper.
- Conceptual clarification: Delimiting and precisifying a vague or overused academic concept. For instance, the concept of "resilience" has been assigned different operational definitions in ecology, psychology, and urban planning — these definitions are, in essence, interpretive specifications of the same term across different contexts.

The common feature of these activities is that they all create new knowledge relations rather than simply reproducing existing knowledge. The interpreter not only states "what the original text said" but, more importantly, makes judgments about "what this means" and "where this can be applied." These judgments constitute the cognitive scaffolding for subsequent researchers — they lower the threshold of understanding, indicate boundaries of applicability, and provide operational pathways.

*3.2 Interpretation vs. Information Extraction: Drawing a Bright-Line Boundary*

A critical cognitive boundary must be drawn here: interpretation is fundamentally distinct from information extraction. Conflating the two would lead to a profound misunderstanding of the IKN concept — mistaking the IKN for a kind of "advanced summary" or "AI-generated text fragment."

Information extraction (including automatic summarization, keyword extraction, text chunking, etc.) answers the question: "What did the original text say?" Its goal is information fidelity — compressing text length while minimizing loss of the original text's informational content. A qualified summary should faithfully present the original text's core findings without adding the summarizer's own judgments.

Interpretation answers the question: "What does this mean? Where can it be applied?" Its goal is sense-making — establishing new connections between existing pieces of knowledge, making judgments about applicability, limitations, and extensibility. The products of these judgments are new knowledge relations, not compressed versions of existing information.

The following example clearly distinguishes the two:

- Original text: The Transformer model achieved 28.4 BLEU on the WMT 2014 English-to-German translation task, surpassing the previous best result [12].
- Information extraction (summary): This paper proposes the Transformer architecture and validates its effectiveness on machine translation benchmarks.
- Interpretation (IKN-level): The Transformer's core contribution lies in replacing recurrent architectures with self-attention, thereby structurally resolving the long-range dependency problem — because any position in the sequence can attend directly to any other, the computational path length between positions is independent of their distance. This structural property makes it especially suited for tasks requiring long-range relationship modeling, such as document-level machine translation and coreference resolution. Its applicability is constrained when computational resources are limited, as self-attention's complexity grows quadratically with sequence length; however, it is the preferred choice in compute-abundant scenarios because its non-sequential computation is far more amenable to parallelization than RNNs.

Note the characteristics of the third text: it is not simply "more detailed" — it makes several judgments that only an interpreter can make. It identifies the Transformer's core contribution (self-attention replacing recurrence), explains the reason for its structural advantage (computational path length independent of sequence position), and derives the applicable scope (long-range relationship modeling tasks) and conditions (compute-constrained scenarios). These judgments require understanding the meaning of the original text, not merely extracting its content.

This distinction is the cornerstone of IKN theory construction. It demonstrates that interpretation is an independent link in the knowledge production process, and its products possess new information irreducible to the original text. If interpretation indeed creates new information — judgments about meaning, applicability, and relatedness — then it deserves to be taken seriously as an independent knowledge object.

*3.3 Why Has Interpretation Long Been Invisible?*

If interpretation is so fundamental and unavoidable in knowledge flow, why has it long lacked an independent scholarly identity? The answer can be attributed to two interrelated historical conditions.

First, subject-dependence. Before the large-scale intervention of AI technologies, high-quality interpretive knowledge was heavily dependent on the cognitive labor of individual experts. A senior researcher's elaboration of a classic theory is the crystallization of decades of reading, thinking, and teaching. Such interpretive knowledge is tightly coupled with the interpreter — it exists in her lectures, her private notes, her informal exchanges with students. Because interpretation was difficult to detach from the interpreter, the scholarly system defaulted to treating interpretation as a competence attached to the person rather than an object independent of the person[13].

Second, lack of an identity mechanism. Even when interpretation was committed to writing — a paragraph in a review article, a chapter in a textbook, a post on an academic blog — it consistently lacked a formal "identity anchor." The review article itself has a DOI, but the interpretation of a specific concept within that review has no independent identifier. The textbook has an ISBN, but the applicability judgment about a particular method within that textbook cannot be independently cited. Interpretation is scattered across various texts, existing in fragmented and subordinate forms, never having acquired an institutional identity that says "this is it"[14].

*3.4 The Structural Consequences of the Interpretation Black Box*

The long-standing lack of objectified identity for interpretive activity is not a neutral condition. It inflicts systematic damage on the efficiency, reliability, and equity of the scholarly knowledge system.

First, reuse entropy increase. Each generation of researchers, when confronted with classic texts, must begin interpretive understanding from scratch. Even though prior generations have already invested enormous cognitive labor in "understanding this text," the fruits of that labor cannot be transmitted to successors in a structured form. The result is large-scale, repetitive investment in understanding — every doctoral student must independently "re-understand" classic theories half a century old from the original texts, because their predecessors' understanding was not objectified, not preserved, not transmitted. This is not because predecessors are unwilling to share, but because no institutional object form exists to carry such sharing.

Second, error propagation. When an erroneous understanding emerges — for example, an overly broad interpretation of the applicability conditions of a statistical method — this error is difficult to trace and correct. Because the interpretation itself lacks an independent identity, there is no anchor point where one can annotate "this interpretation has been superseded by a more precise understanding." Erroneous interpretations can propagate continuously through word of mouth, informal training, and vague textual citation, while the scholarly system lacks a mechanism for flagging and correcting such propagation.

Third, contribution invisibility. Researchers who primarily make interpretive contributions — scholars who write widely used textbooks, mentors who influence an entire generation of researchers through their teaching, authors of methodological guides that enable the dissemination of complex techniques — find their contributions difficult to measure accurately within the current evaluation system. A textbook may be cited tens of thousands of times, yet these citations are assigned far less weight in impact factor calculations than original research papers. This is not a problem with citation metrics per se — it is because the scholarly object system contains no measurement category for "interpretive contribution." Interpretive work cannot enter the evaluation system, not because the evaluation system is biased, but because the object form that would make such evaluation possible is absent.

## 4 From Node to Object: Defining the IKN

The preceding three sections have established the necessity of the IKN concept: the paper as a scholarly object has insurmountable boundaries (Section 2), and interpretation, though indispensable in knowledge flow, lacks object identity (Section 3). The task of this section is positive construction: What is an IKN? Why is it both a "node" and an "object"?

### *4.1 Defining the Interpretive Knowledge Node (IKN)*

Based on the foregoing argument, this paper offers the following complete definition of the IKN:

*An Interpretive Knowledge Node (IKN) is an interpretive knowledge object that is based on an explicit scholarly source, formed through the structured interpretation of an existing knowledge unit and certified by a scholarly subject, possessing an independent identity, and capable of being tracked and cited.*

Each qualifier in this definition carries a distinct theoretical intention:

- "Based on an explicit scholarly source": An IKN is not a knowledge claim conjured from nowhere but is anchored to a specific knowledge unit within an existing scholarly document. It must include a precise source pointer to the interpreted object — a DOI, chapter, paragraph, figure/table number, or data location. This requirement distinguishes the IKN from general "knowledge claims."
- "Structured interpretation": The core content of an IKN is not a reiteration or summary of the interpreted object but a structured exposition of its meaning, applicability conditions, boundaries, and relationships with other knowledge units. It answers "what does this mean" and "where can this be applied."
- "Certified by a scholarly subject": The object identity of an IKN is not automatically acquired. An interpretation must undergo certification by a member of the scholarly community (the interpreter or another qualified subject) to obtain the formal identity of an IKN. This requirement distinguishes the IKN from AI-generated text fragments and from unexamined personal understandings.
- "Independent identity": An IKN possesses its own unique identifier, independent of both the source object it interprets and the interpreter who created it.
- "Capable of being tracked and cited": An IKN should be locatable within the knowledge network and serve as a formal evidentiary basis in scholarly writing. During actual invocation, subsequent literature directly embeds the IKN pointer, allowing the system to dynamically fetch and render its underlying structured interpretation at the reading or compilation interface, thereby achieving seamless reuse and precise traceability.

*4.2 Why Both "Node" and "Object"?*

The name "IKN" contains a seemingly paradoxical formulation: it is both a "Node" — implying a position within a network — and something we argue should be an "Object" — implying an independent entity. Resolving this dual identity is key to understanding the theoretical positioning of the IKN.

The node dimension emphasizes the IKN's structural function. In the citation network analysis framework of the first paper[9], the IKN was identified as an intermediate node situated between source documents and citing documents. It receives knowledge from upstream sources and transmits organized and standardized knowledge content to downstream citers. In the graph's topological structure, the IKN exhibits high betweenness centrality — a large number of citation paths flow through it, connecting classics to frontiers. This structural function is the external manifestation of the IKN observable within the knowledge network.

The object dimension emphasizes the IKN's object identity. It is not merely a position through which things flow — it is itself an entity with an independent reason for existence. Just as a train station is not merely a point that is passed through but a facility with independent architectural structure and institutional function, the IKN is not merely a waystation on a citation path; it carries content irreducible to pathway function — namely, interpretive knowledge that has been structurally organized and communally certified.

Thus, the IKN is a scholarly object that exists in the form of a node. "Node" describes its position and function within the knowledge network; "Object" asserts its independent existential qualification. The two are not opposing concepts but descriptions of the same thing from different analytical dimensions: from the network analysis dimension, the IKN is a node; from the knowledge object theory dimension, the IKN is an object.

*4.3 The Three Core Dimensions of the IKN*

Having provided a complete definition and clarified the dual identity, we can distill the three core dimensions that together constitute the analytical skeleton of the IKN concept:

Source-Anchoring (Provenance). An IKN must be explicitly bound to a specific scholarly source object. This dimension ensures that the IKN has an objective, verifiable basis — any subsequent researcher can trace back to the interpreted original text and independently judge whether the IKN's interpretation is faithful to the source. Source-anchoring is the fundamental distinction between an IKN and a general "knowledge claim" or "personal opinion."

Interpretation Fixation. An IKN records a stable "state of understanding" that has been organized and certified, not a dynamic, private thought process. This dimension ensures that the IKN possesses the temporal stability required of a scholarly object — once an IKN is certified, its core interpretive content is fixed (subsequent modifications will produce new versions through a version management mechanism rather than modifying the existing version).

Functional Reusability. The core purpose of an IKN is to provide directly mobilizable interpretive judgments for subsequent research. A researcher confronting a classic text can consult existing IKNs to quickly grasp its core contributions, applicability conditions, and limitations — without having to begin independent interpretive labor from the original text. This dimension defines the "use value" of the IKN within the scholarly ecosystem.

## 5 The Four Attributes of the IKN as a Scholarly Object

Section 4 provided the definition and core dimensions of the IKN. The task of this section is to go further: to systematically argue that the IKN satisfies the qualifying conditions of a "scholarly object." We propose a Four-Attribute Qualification Framework — an entity must simultaneously satisfy the four conditions of provenance, interpretability, human recognition, and citability to qualify as an IKN. This framework not only provides criteria for judging the object identity of the IKN but also offers a referable qualification template for other new scholarly objects that may emerge in the future. The following argues, attribute by attribute, that the IKN satisfies each criterion.

### *5.1 Provenance: Source-Anchoring*

A scholarly object must possess a clear, verifiable source. This is not a pedantic formal requirement but an essential feature distinguishing scholarly knowledge from other types of knowledge: scholarly knowledge is "scholarly" precisely because each of its assertions can be traced back to its evidentiary basis. A paper must declare its data sources and cited literature; a dataset must explain its collection methodology and processing pipeline; software must provide its source code or algorithmic description. The transparency and traceability of sources is the first attribute of a scholarly object.

For the IKN, provenance means it must clearly answer one question: "What does it interpret?" Concretely, an IKN must include precise source pointers to the interpreted object, including but not limited to: the DOI of the target paper, the number or title of the target section, the in-document location of the target equation or figure/table, or the persistent identifier of the target dataset. This requirement generates two constraints:

- Positive constraint: The IKN's interpretive content must be verifiable against the original text. Any interpretation that claims to be based on a particular paper must enable a third party to locate the relevant portion of that paper and independently judge the interpretation's fidelity and accuracy.
- Negative constraint: Knowledge assertions that cannot be anchored to an explicit scholarly source do not qualify as IKNs. A vague generalization about "what the academic community generally believes" or a loose statement about "the consensus in a field" lacks the source-anchoring that an IKN requires.

Provenance provides the IKN with an objective basis for objectification: it is no longer a vague opinion floating in the space of scholarly discourse but a concrete knowledge object rooted in locatable, verifiable scholarly literature.

### *5.2 Interpretability*

Where provenance answers "where does the IKN come from," interpretability answers "what does the IKN provide."

An entry that merely provides a source pointer without any interpretive content — for example, "Paper X proposed Concept Y" — does not suffice as an IKN. At best, it is a citation entry or index record. The core value of an IKN lies in the interpretive judgments it carries: structured expositions of the interpreted object's meaning, applicability conditions, boundaries, and relatedness. These judgments constitute knowledge that "the original text itself does not contain but is essential for the effective use of that text."

The content of interpretability can be further decomposed into three levels:

- Meaning attribution: Where does this knowledge unit sit within the knowledge structure of its domain? What problem does it solve? What is its core contribution? The answers to these questions are not always directly extractable from the original text — they often require the interpreter to position the original text within a larger knowledge graph in order to make such judgments.
- Applicability judgment: Under what conditions is this knowledge unit applicable? Under what conditions is it not? What are its theoretical premises or experimental assumptions? These judgments are crucial for subsequent researchers to correctly use the knowledge unit, yet they too require reasoning that transcends the original text.
- Relation construction: To which other knowledge units is this knowledge unit related? What does it inherit? By what has it been superseded? With what does it exist in tension? This level of interpretation effectively establishes traversable network connections between knowledge units.

It must be especially noted that interpretive judgments are not subjective speculation. They are subject to dual constraints: provenance (the interpretation must be faithful to the interpreted object) and human recognition (see 5.3). An IKN's interpretation can be corrected, updated, or even replaced in subsequent use — but this process of correction is itself a new form of knowledge accumulation: it not only corrects "our understanding of a certain text," but this correction itself is recorded and tracked.

*5.3 Human Recognition*

This attribute is the key distinguishing the IKN from AI-generated content.

Within the framework of this paper, which defines the IKN as a "scholarly object," an interpretive knowledge unit must undergo human recognition by the scholarly community to acquire object identity. The logic here is not "whoever creates, owns," but rather "whoever uses, endorses" — the legitimacy and credibility of an IKN derive not from the declaration of its creator but from the repeated certification it receives through the citing and signing behaviors of subsequent users.

The logic of human recognition is continuous with the peer review mechanism for papers. A paper is regarded as "qualified scholarly knowledge" not because it was written but because it passed peer review — i.e., it underwent the scrutiny and endorsement of other qualified subjects within the community. The IKN extends this logic to interpretive knowledge: an interpretation becomes a qualified scholarly object because it has undergone the scrutiny, use, and endorsement of other researchers — not merely because someone wrote it down.

A critical logical distinction must be emphasized: human recognition is a qualifying condition for the IKN as an object, not a necessary condition for the IKN to exist. An uncertified interpretation still exists, and may still be useful, but it lacks the institutional guarantee that makes something a "scholarly object." Just as an un-peer-reviewed manuscript still conveys knowledge but is not regarded as a "paper" with formal object status, an uncertified IK candidate is an "interpretation manuscript" — it has value but lacks the identity of a formal scholarly object.

This attribute also lays the theoretical foundation for the third paper in the series — concerning the production mechanism combining AI-assisted generation with human certification. AI can assist in producing candidate interpretations, but the "certification" step that confers scholarly object identity upon an interpretation must be performed by the scholarly community.

*5.4 Citability*

Citability is the ultimate hallmark of becoming a scholarly object. It means that the object is no longer merely background information or an informal reference in scholarly communication but can formally appear in the reference list of scholarly literature as an independent link in the chain of argument.

The act of citation carries multiple functions in academia: it traces the origin of ideas, allocates scholarly credit, embeds arguments within intellectual genealogies, and provides subsequent researchers with traceable retrieval pathways[15]. When an object possesses citability, it enters the credit economy of the scholarly system — it can gain citations (be recognized) and can lose citations (be forgotten or replaced). This entry marks an object's upgrade from "existence" to "institutional existence."

For the IKN, citability means: when a researcher writing a paper uses the interpretive judgment provided by a certified IKN (for example, regarding the applicability conditions of a particular method), she can directly cite that IKN in her reference list, just as she would cite an original paper or a dataset. The institutional implications of this act are profound: it means that interpretation itself, for the first time, becomes a formal evidentiary basis.

Before this, "my understanding of a certain paper" could appear in scholarly writing in only two ways: either as the author's own argument (no source cited, because the "understander" is the author herself), or as a citation to a review article (citing the review article as a whole, not the specific interpretation of a particular concept within it). The citability of the IKN makes a third way possible: citing an independent, certified interpretive knowledge object as a formal evidentiary basis for an argument.

## 6 Systematic Comparison of the IKN with Existing Scholarly Objects

The preceding sections have completed the positive argument for the IKN as a scholarly object. This section further clarifies the IKN's unique position within the scholarly object system through systematic comparison. The following table situates the IKN within the spectrum of existing scholarly objects:

| Scholarly Object | Core Task | Objectifies "Interpretation"? | Granularity | Independent Citation |
|---|---|---|---|---|
| Paper | Records research outputs | No — interpretation embedded within argumentative text | Article-level | ✓ |
| Dataset | Provides data materials | No | Data-level | ✓ |
| Software | Implements algorithms or computational functions | No | Code-level | ✓ |
| Review Article | Synthesizes domain progress | Partially, but as the author's personal viewpoint, not as an | Domain-level | ✓ |

| Scholarly Object | Core Task | Objectifies "Interpretation"? | Granularity | Independent Citation |
|---|---|---|---|---|
| | | independent object | | |
| Annotation | Marks text fragments | No — dependent on the marked text | Fragment-level | ✗ |
| Nanopublication | Expresses a single verifiable assertion | No — focuses on the assertion itself, not interpretation | Assertion-level | ✓ |
| IKN | Solidifies certified interpretive knowledge | Yes — the first proposal to formalize the interpretive knowledge unit as an independent scholarly object | Knowledge-unit-level | ✓ |

This comparison reveals a clear line of differentiation: the primary function of existing scholarly objects is preservation or expression — preserving research outputs (Paper), preserving data (Dataset), preserving code (Software), preserving a bird's-eye view of a research domain (Review Article), preserving textual markup (Annotation), preserving atomized assertions (Nanopublication). None of them is specifically designed for preserving interpretation.

The Review Article comes closest to the IKN, but a key difference remains: the interpretation within a review is the author's personal scholarly viewpoint, not an independent citable object. When you cite a review, you cite "a review paper published by Author X in Year Y," not "a specific interpretation of a particular concept within that review." Interpretation is encapsulated within the review as a whole, lacking an independent path for extraction and citation.

The distinction between Annotation and the IKN lies in the question of independence. Annotations (whether Web Annotations, Hypothesis.is annotations, or other forms of textual markup) are technically dependent on the annotated text — the annotation's raison d'être derives from its pointing to a textual location. The IKN, though also anchored to a source document, uses anchoring as a means of establishing credibility rather than as the purpose of its existence. The IKN's independent object identity means it can be independently accessed, independently cited, and independently discussed without directly displaying the full text of the source document. An analogy: a paper cites a reference; the paper cannot exist independently of this citation relationship (it is embedded in an intellectual genealogy), but no one would say that the paper is "dependent on" the reference. The IKN's relationship to its source document is analogous — the anchoring relationship constitutes part of the IKN's credibility, but the IKN itself is an independent object.

The distinction between Nanopublication[16] and the IKN lies in cognitive content. The Nanopublication aims to express a "minimal publishable assertion" — typically "Subject S asserts P," accompanied by provenance and metadata. Its core goal is to enable machine-readable atomized facts to enter knowledge graphs. But the Nanopublication focuses on machine-readable assertions rather than contextualized interpretation: it tells you "someone claims X," but its design goals do not include judgments about "what X means" or "under what conditions X applies." The IKN's core is precisely this interpretive dimension — "not only telling you what something is, but telling you what it means."

In summary, existing objects preserve "documents" or "data" or "assertions." The IKN proposes a method for formalizing "certified interpretive knowledge units" as independent scholarly objects, filling a structural gap in the scholarly object system.

## 7 Discussion: Boundaries and Pathways

*7.1 The IKN Does Not Replace the Paper*

When proposing a new scholarly object, the easiest misunderstanding to arise is viewing it as a replacement for or competitor to the existing dominant object. This paper must clarify unequivocally: The IKN does not replace the paper, nor does it compete with the paper.

The division of labor between the two is clear. The paper records the research process — how the researcher posed the question, designed the method, collected and analyzed data, and reached conclusions. The paper's contribution lies in producing new knowledge. The IKN records interpretive results — what existing knowledge means, under what conditions it applies, and what relations it bears to other knowledge. The IKN's contribution lies in providing structured pathways of understanding for existing knowledge.

This relationship can be analogized to that between a map and its landmarks. Papers are landmarks — they establish new discoveries in the knowledge space. IKNs are the annotations on the map — they explain what a landmark is, how to reach it, and its relationship to surrounding landmarks. Without landmarks, the map is meaningless; without annotations, landmarks are hard to find and use. The two are not in a competitive relationship but a complementary one.

*7.2 The IKN Is an Object, Not Merely an AI Product*

Against the backdrop of rapid AI technological development, it is tempting to equate the IKN with a kind of "AI-generated summary" or "large model output." This paper's position is unequivocal: The legitimacy and identity of the IKN derive from the certification and objectification processes of the scholarly community, not from the generative capacity of AI.

It is crucial to distinguish "possible modes of IKN production" from "the object identity of the IKN." AI can play an important role in the IKN production pipeline — multi-agent systems can generate multi-perspective candidate interpretations of the same knowledge source, and cross-validation can reduce the biases of single models — these technical possibilities will be explored in detail in the third paper of the series. But regardless of what role AI plays in producing candidate IKNs, the "certification" step that confers scholarly object identity upon an IKN must be performed by the scholarly community.

This is entirely consistent with the logic of papers: AI can assist in writing papers (grammar checking, reference formatting, even draft generation), but a paper becomes a "paper" because it has passed through peer review and formal publication processes — these are institutional acts of the human scholarly community. The IKN follows the same logic: AI can help produce candidate interpretations, but an interpretive knowledge unit becomes an IKN because it has undergone certification and adoption by scholarly subjects.

*7.3 Governance Challenges*

This paper is a concept-construction paper; its task is to propose and argue for the IKN's qualification as a new scholarly object, not to resolve all the problems of its implementation. However, the following governance challenges are worth identifying even at the stage of theoretical construction, as an agenda for subsequent research:

Version management. As scholarly understanding deepens, the interpretation carried by an IKN may need updating — for example, a judgment about the applicability conditions of a method may require revision in light of new experimental results. The IKN requires a version management mechanism that enables the evolution of interpretation to be tracked (with old versions preserved as an archive of "historical understanding") while always allowing users to locate the latest certified version. This requirement bears some analogy to software version management (such as Git) and dataset version management (such as Dataverse), but it needs to be adapted to the characteristics of scholarly knowledge.

Dispute resolution. For a given knowledge unit, multiple competing or even contradictory interpretations may exist. In the paper system, such disputes are handled through the publication of multiple papers, scholarly debate, and consensus formation. The IKN system requires a similar but potentially more structured dispute resolution mechanism: should multiple competing IKNs be allowed to coexist (analogous to multiple papers holding different views), or should some consensus mechanism converge toward a single certified version? These questions exceed the scope of this paper but need serious attention in institutional design.

Certification process. Regarding the question of "who is qualified to certify an IKN," there exists a continuum from permissive to strict. The most permissive approach is "any citing party can perform implicit certification through the act of citation" (analogous to how citing a paper itself constitutes a degree of endorsement). The strictest is "only reviewers with specific qualifications can grant formal IKN certification." Different certification schemes will produce different IKN ecosystems: too permissive may lead to uneven IKN quality; too strict may recapitulate the protracted timelines of paper publication. The design of the certification process is one of the most central institutional design questions in moving the IKN from concept to realization.

## 8 Conclusion

The modern scholarly system has accomplished a remarkable institutional engineering feat: it has successfully objectified the "products of knowledge production." Papers, datasets, software — these scholarly objects enable human knowledge-creation activities to be recorded, disseminated, cited, and accumulated. That science has been able to advance with unprecedented speed and technical precision is, to a considerable extent, the result of the effective operation of this scholarly object system.

Yet a structural void exists within this system.

Between knowledge production and knowledge reuse lies a critical but long-overlooked link: interpretation. Every researcher, when using existing knowledge, must transform it into an understandable and mobilizable form — a transformation that is not merely textual reiteration but involves meaning attribution, applicability judgment, and relation construction. These interpretive activities, though they constitute the factual foundation of knowledge flow, have never acquired an independent scholarly identity. They are scattered across the prose of papers, the chapters of textbooks, the lectures of classrooms — omnipresent yet nowhere to be found.

The Interpretive Knowledge Node (IKN) proposed in this paper aims to fill this structural void. By formalizing interpretive activity as a scholarly object possessing the four attributes of provenance, interpretation fixation, human recognition, and citability, the IKN endows "interpretation" with an independent scholarly identity. It does not replace the paper — the paper remains the optimal object for recording research outputs. What it does is expand the boundary of the scholarly object system, enabling the scholarly system to record not only "what scientists discovered" but also "how scientists understand those discoveries."

The significance of this expansion transcends the addition of a term or concept. If the IKN is institutionalized, the efficiency of scholarly knowledge dissemination will be enhanced by the objectification of interpretation — each generation of researchers will not need to independently understand the classics from scratch; they can advance by standing on the understanding of their predecessors. If the IKN is institutionalized, interpretive contributions will receive measurable scholarly recognition — producing a compelling interpretation will, like making an original discovery, become a scholarly act that can be cited, tracked, and evaluated. If the IKN is institutionalized, erroneous understandings can be flagged, tracked, and corrected — knowledge dissemination will be not only faster but also more reliable.

Of course, a vast distance separates conceptual construction from institutional realization. The IKN's version management, certification processes, dispute resolution, and integration with existing publishing and indexing infrastructures — these governance and technical problems must be resolved one by one in subsequent research. The contribution of this paper is to provide the conceptual foundation for that subsequent work: it has argued why the IKN should exist. As for how to realize it, that is the task of the subsequent papers in this series.

## References

1. Fischhoff B. The sciences of science communication[J]. Proceedings of the National Academy of Sciences, 2013, 110(supplement_3): 14033-14039. DOI: 10.1073/pnas.1213273110.
2. Borgman C L, Furner J. Scholarly communication and bibliometrics[J]. Annual Review of Information Science and Technology, 2002, 36(1): 2-72. DOI: 10.1002/aris.1440360102.
3. Latour B, Woolgar S. Laboratory Life: The Construction of Scientific Facts[M]. Princeton: Princeton University Press, 1986.
4. Paskin N. Digital Object Identifier (DOI®) System[M]//Bates M J, Maack M N, eds. Encyclopedia of Library and Information Sciences. 3rd ed. Boca Raton: CRC Press, 2010: 1586-1592.
5. Garfield E. Citation analysis as a tool in journal evaluation: Journals can be ranked by frequency and impact of citations for science policy studies[J]. Science, 1972, 178(4060): 471-479. DOI: 10.1126/science.178.4060.471.
6. Van de Sompel H, Payette S, Erickson J, et al. Rethinking Scholarly Communication[J]. D-Lib Magazine, 2004, 10(9): 1082-9873.
7. Garfield E. Citation indexes for science: A new dimension in documentation through association of ideas[J]. Science, 1955, 122(3159): 108-111. DOI: 10.1126/science.122.3159.108.
8. McGrath W. The unit of analysis (objects of study) in bibliometrics and scientometrics[J]. Scientometrics, 1996, 35(2): 257-264. DOI: 10.1007/bf02018483.
9. Li L, Cao Y. Citation Pathways in the AI Era: Interpretive Knowledge Nodes, Citation Compression Layers, and the Measurement Boundary of Scholarly Impact[J]. arXiv preprint arXiv:2607.18350, 2026. DOI: 10.48550/arXiv.2607.18350.
10. Royal Society. History of Philosophical Transactions[EB/OL]. (2015-03-06)[2026-07-30]. https://royalsociety.org/journals/publishing-activities/publishing350/history-philosophical-transactions.
11. Burnham J C. The evolution of editorial peer review[J]. JAMA, 1990, 263(10): 1323-1329. DOI: 10.1001/jama.1990.03440100023003.

12. Vaswani A, Shazeer N, Parmar N, et al. Attention Is All You Need[C]//Advances in Neural Information Processing Systems 30. Long Beach: Neural Information Processing Systems Foundation, 2017: 5998-6008.
13. Venkitachalam K. Tacit knowledge: Review and possible research directions[J]. Journal of Knowledge Management, 2012, 16(2): 357-372. DOI: 10.1108/13673271211218915.
14. De Smedt K, Koureas D, Wittenburg P. FAIR digital objects for science: From data pieces to actionable knowledge units[J]. Publications, 2020, 8(2): 21. DOI: 10.3390/publications8020021.
15. Price D J D S. Networks of scientific papers: The pattern of bibliographic references indicates the nature of the scientific research front[J]. Science, 1965, 149(3683): 510-515. DOI: 10.1126/science.149.3683.510.
16. Mons B, van Haagen H, Chichester C, et al. The value of data[J]. Nature Genetics, 2011, 43(4): 281-283. DOI: 10.1038/ng0411-281.